\documentclass{PoS}

\usepackage{graphics,graphicx}
\usepackage{multicol} 
\usepackage{parskip}
\usepackage{amsmath}
\usepackage{multirow}
\usepackage[utf8]{inputenc}
\usepackage{fancyhdr}
\usepackage[title]{appendix}
\usepackage{wasysym}
\usepackage{url}

\title{Core-Collapse Supernove Burst Neutrinos in DUNE}

\ShortTitle{Core-Collapse Supernove Burst Neutrinos in DUNE}

\author{\speaker{C. Cuesta} on behalf of DUNE collaboration \\
        Centro de Investigaciones Energ$´{e}$ticas, Medioambientales y Tecnol$´{o}$gicas,
 CIEMAT, \\ 28040, Madrid, Spain\\
        E-mail: \email{clara.cuesta@ciemat.es}}

\abstract{The Deep Underground Neutrino Experiment (DUNE), a 40-kton fiducial mass underground liquid argon time projection chamber experiment, will be sensitive to the electron-neutrino-flavor component of the burst of neutrinos expected from the next Galactic core-collapse supernova. Such an observation will bring unique insight into the astrophysics of core collapse as well as into the properties of neutrinos. The recent progress on detection and reconstruction of supernova burst neutrinos in DUNE, including the contribution of the light detection systems are presented.}

\FullConference{%
  40th International Conference on High Energy physics - ICHEP2020\\
  July 28 - August 6, 2020\\
  Prague, Czech Republic (virtual meeting)
}

\begin{document}

\section{Supernova neutrino burst}

Core-collapse supernovae (SN) occur when a massive star dyes and are a huge source of neutrinos of all flavors in our Universe. During a SN explosion, 99\% of the gravitational binding energy of the star ($\sim10^{53}$~ergs) is released by neutrinos and antineutrinos of all flavors, which play the role of astrophysical messengers, escaping from the SN core. 

\begin{figure}[htp]
    \centering
    \includegraphics[width=0.7\textwidth]{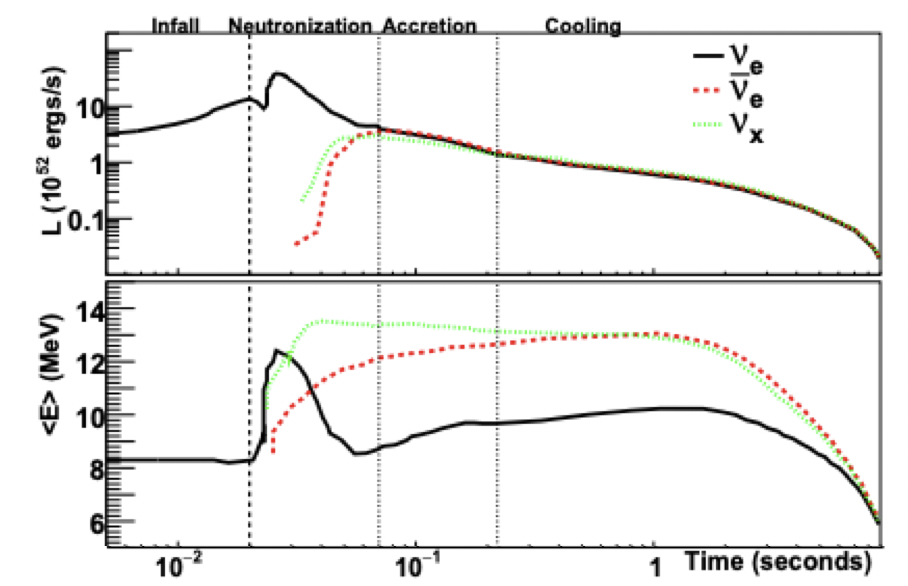}
    \caption{Expected time-dependent flux parameters for a specific model for an electron-capture supernova~\cite{Huedepohl:2009wh}. No flavor transitions are assumed. The top plot shows the luminosity as a function of time, and the bottom plot shows average neutrino energy.}
    \label{fig:1}
    \end{figure}

We can tell the story of the SN looking at the neutrino emission. SN neutrinos are emitted in a burst of a few tens of seconds duration~\cite{Huedepohl:2009wh}. Essentially three stages can be distinguished ,as shown in Figure~\ref{fig:1}: the neutronization burst, where a large $\nu_e$ emission takes place in the first 10's of ms; the accretion phase, which lasts between few tens to few hundreds of ms, and where the luminosity of electron and non-electron flavors is significantly different; and the cooling phase, up to $\sim$10~s, when luminosities and average energies decrease and there is almost luminosity equipartition between neutrino species.

The measurement of the neutrino energy spectra, flavor composition and time distributions from SN will provide critical information about the SN dynamics and neutrino properties as neutrino mass ordering can be obtained.

\section{The DUNE far detector}

The DUNE experiment aims to address key questions in neutrino physics and astroparticle physics \cite{tdr_v1, lbl, sn, bsm}. It includes precision measurements of the parameters that govern neutrino oscillations with the goal of measuring the CP violating phase and the neutrino mass hierarchy with a muon neutrino beam produced at Fermilab. The physics programme also addresses non-beam physics as nucleon decay searches and the detection and measurement of the electron neutrino flux from a core-collapse supernova within our galaxy. The ancillary program for underground physics in DUNE also includes the measurement of neutrino oscillations using atmospheric neutrinos, solar neutrinos, detecting diffuse supernova neutrino fluxes and searches for neutrinos from extra-solar astrophysical sources.

DUNE will consist of a near detector placed at Fermilab close to the production point of the muon neutrino beam of the Long-Baseline Neutrino Facility (LBNF), and four 10\,kt fiducial mass liquid argon time projection chambers (LAr TPCs) as far detector in the Sanford Underground Research Facility (SURF) at 4300\,m.w.e. depth at 1300\,km from Fermilab. 

SN neutrinos will be detected by the DUNE far detector, composed by four modules of 70-kton liquid argon mass in total of which 40 kton will be fiducial mass. LArTPC technology provides good energy resolution and full particle reconstruction with very high quality tracking. Energy thresholds as low as a few MeV may be possible. In these detectors, the ionization charge is drifted by an electric field towards the anode where the charge is collected. Using the time arrival of the charge at the readout planes, a three-dimensional track reconstruction is possible.  Particles are identified by the rate of energy loss along the track.  The Ar scintillation light is also detected enabling fast timing of signals and event localization inside the detector. Different LAr technologies are being considered for the DUNE far detector: Single-phase (SP) LArTPC technology is designed to have horizontal drift of 3.5 m with wrapped-wire readout including two induction and one charge collection anode planes \cite{tdr_v4}; and dual-phase (DP) LArTPC technology has vertical drift over 12 meters, and at the liquid-gas interface at the top of a DP module, drifted ionization charge is amplified and collected \cite{idr_v3}.

\section{Low energy events in DUNE}

\begin{figure}[htp]
    \centering
    \includegraphics[width=0.6\textwidth]{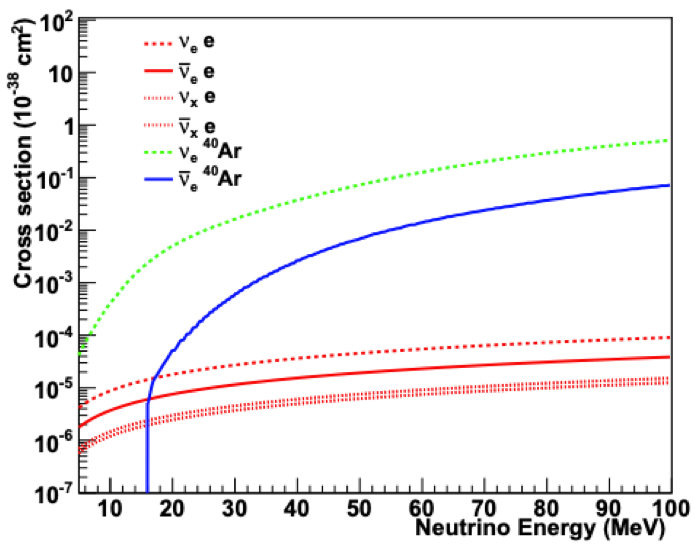}
    \caption{Cross sections for supernova-relevant interactions in argon as a function of neutrino  energy.}
    \label{fig:2}
    \end{figure}

There are different detection channels in LAr for the neutrino interactions. The dominant interaction is the charged-current absorption of $\nu_e$ on $^{40}Ar$, for which the observable is the e$^-$ plus deexcitation products from the excited $^{40}K^{\ast}$ final state. Additional channels include a $\bar{\nu}_e$ CC interaction and electron scattering. Cross sections for the most relevant interactions are shown in Fig.~\ref{fig:2}. The predicted event rate from a supernova burst may be calculated by folding expected neutrino flux differential energy spectra with cross sections for the relevant channels, and with detector response; this is done using SNOwGLoBES \cite{snowglobes}. 

\begin{figure}[htp]
    \centering
    \includegraphics[width=1\textwidth]{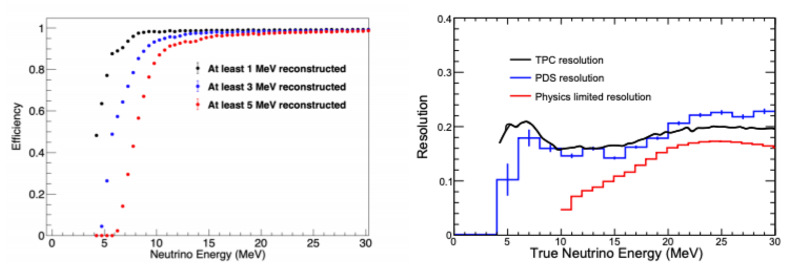}
    \caption{Left: reconstruction efficiency as a function of neutrino energy for MARLEY $\nu_e$CC events, for different minimum
required reconstructed energy. Right: fractional energy resolution as a function of neutrino energy for TPC tracks corrected for drift
attenuation (black) and photon detector calorimetry (blue)}
    \label{fig:3}
    \end{figure}

We use MARLEY \cite{marley} to simulate tens-of-MeV neutrino-nucleus interactions in liquid argon and the LArSoft \cite{larsoft} Geant4-based software package to simulate the final-state products from MARLEY in the DUNE LArTPC. SN neutrino events due to their low energies manifest as spatially small events of few centimeters. Figure \ref{fig:3} shows summarized fractional energy resolution and efficiency performance for MARLEY events. In addition, understanding of cosmogenic and radiological backgrounds, dominated by $^ {39}$Ar, is necessary, although we expect a minor impact on reconstruction, the triggering efficiency could be affected.

\begin{figure}[htp]
    \centering
    \includegraphics[width=0.7\textwidth]{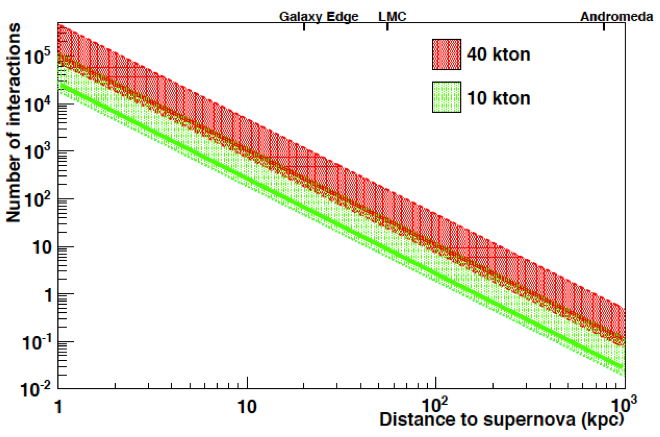}
    \caption{Estimated numbers of supernova neutrino interactions in DUNE as a function of distance to the supernova, for different detector masses.}
    \label{fig:4}
    \end{figure}

With these tools, we estimate the rates calculated for the dominant interactions. The results of these studies are detailed in~\cite{sn}. Clearly $\nu_e$ dominate and LAr is the only future prospect for a large, cleanly tagged SN sample. Figure \ref{fig:4} shows expected SN neutrino interactions as a function of the distance to the SN. Core collapses are expected to occur a few times per century, at a most-likely distance of around 10 to 15 kpc. For a collapse in the Andromeda Galaxy, almost 800 kpc away, a 40 ton detector will observe a few events.

\begin{figure}[htp]
    \centering
    \includegraphics[width=0.55\textwidth]{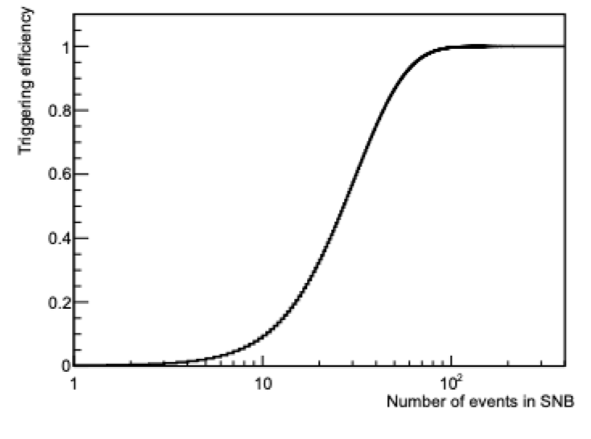}
    \caption{Supernova neutrino burst triggering efficiency for the DP photon detectors as a function of the number of interactions in one module.}
    \label{fig:5}
    \end{figure}

In DUNE, the trigger on a supernova neutrino burst can be done using either TPC or photon detection system information. In both cases, the trigger scheme exploits the time coincidence of multiple signals over a timescale matching the supernova luminosity evolution. A redundant and highly efficient triggering scheme is under development. We describe here an example of a preliminary trigger design study based on the photon detection system of the DP module where the real time algorithm provides the trigger primitives by searching for PMT hits and optical clusters based on time and spatial information. The triggering efficiency as a function of the number of supernova neutrino interactions is shown in Figure \ref{fig:5}. At 20 kpc, the edge of the Galaxy, about 80 supernova neutrino interactions in one module are expected, as it is showed in Figure~\ref{fig:4}. Therefore, the DP photon detection system should yield a highly efficient trigger for a supernova neutrino burst occurring anywhere in the Milky Way.

\section{What can we learn from SN neutrino detection in DUNE?}

A number of astrophysical phenomena associated with supernovae are expected to be observable in the supernova neutrino signal, providing a remarkable  window into the event. In particular, the supernova explosion mechanism, which in the current paradigm involves energy deposition into the stellar envelope via neutrino interactions, is still not well understood, and the neutrinos themselves will bring the insight needed to confirm or refute the paradigm. There are many other examples of astrophysical observables, more details can be found in \cite{sn}.

We have investigated how well it will be possible to fit to the supernova spectral parameters. The physics of neutrino decoupling and the spectra formation is not trivial, but can be parameterized at a given moment by this equation:
\begin{equation}
    \phi (E_\nu) = N \left (\frac{E_\nu}{\langle E_\nu \rangle }\right )^\alpha \exp{\left [-(\alpha+1) \frac{E_\nu}{\langle E_\nu \rangle } \right ]}
\end{equation}

\begin{figure}[htp]
    \centering
    \includegraphics[width=0.7\textwidth]{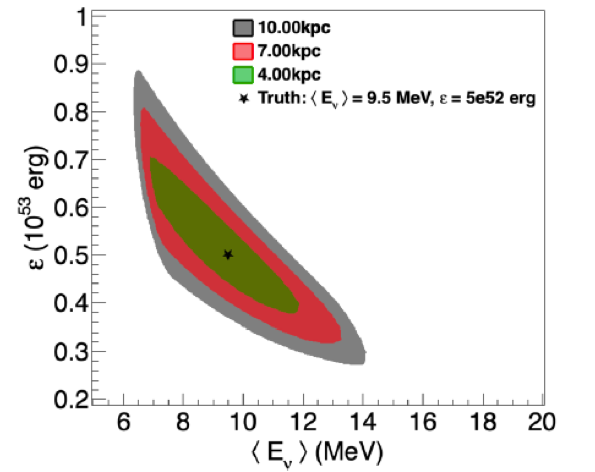}
    \caption{Sensitivity regions for three different supernova distances (90\% C.L.).}
    \label{fig:6}
    \end{figure}

It depends on the neutrino energy, the mean neutrino energy, a "pinching parameter" ($\alpha$), and a normalization constant related to the total luminosity. The different flavors have different average energies and "pinching parameter" and evolve differently in time. We apply a fitting algorithm at a SNOwGLoBES-generated energy spectrum for a SN at a given distance and a chosen true set of pinched-thermal parameters. Figure \ref{fig:6} shows the precision with which DUNE can measure two of the spectral parameters: the binding energy, $\epsilon$, and the average energy of the electron neutrino component for the time-integrated spectrum profiling over $\alpha$.

Detecting neutrinos from a SN, we can also learn a lot about neutrinos and particle physics. A SN can be thought as an extremely hermetic system, which can be used to search for new new physics like Goldstone bosons, neutrino magnetic moments, "dark photons", "unparticles", extra-dimensional gauge bosons, and  sterile neutrinos. Also, self-interactions of neutrinos, neutrino instability, and light gauge bosons can be studied. Such energy-loss-based analysis will make use of two types of information. First, the total energy of the emitted neutrinos compared with the expected release in the gravitational collapse. Second, the rate of cooling should be measured and compared with what is expected from diffusion of the standard neutrinos. As DUNE is mostly sensitive to  $\nu_e$, complementary data of  $\bar{\nu}_e$ from water Cherenkov and scintillator experiments for careful analysis of the flavor transition will be very useful. The flavor oscillation  neutrino physics and its signatures are a major part of the physics program in the different periods.

\section{Conclusions}

The DUNE experiment will be sensitive to neutrinos with about 5 MeV up to several tens of MeV, the regime of relevance for core-collapse supernova burst neutrinos.  This low-energy regime presents particular challenges for triggering and reconstruction. Backgrounds are expected to be relatively tractable during the few tens of second interval of a supernova burst. DUNE's TPC and photon detection systems will both provide information about these events, and we have developed software tools that enable preliminary physics and astrophysics sensitivity studies. DUNE will have good sensitivity to the entire Milky Way, and possibly beyond, depending on the neutrino luminosity of the core-collapse supernova. We expect good sensitivity to supernova electron neutrino spectral parameters. The observation of a burst will also enable sensitivity to neutrino mass ordering, collective effects, and potentially many other topics.

\section*{Acknowledgments}

This project has received funding from the European Union Horizon~2020 Research and Innovation programme under Grant Agreement no.~654168; from the Spanish Ministerio de Economia y Competitividad (SEIDI-MINECO) under Grants no.~FPA2016-77347-C2-1-P, and MdM-2015-0509; and from the Comunidad de Madrid.

\bibliographystyle{JHEP}
\bibliography{biblio}

\end{document}